\documentclass[aps,prd,reprint,onecolumn,11pt,notitlepage,tightenlines,superscriptaddress,nofootinbib,amsmath,amssymb]{revtex4-1}
\pdfoutput=1

\usepackage{graphicx}
\usepackage{microtype}
\usepackage{xurl}
\usepackage[colorlinks,allcolors=blue,pdfusetitle]{hyperref}
\hypersetup{pdfauthor={Changjie Dai, Linyun He, Yonghua Wang, and Wei Chao}}
\makeatletter
\AtBeginDocument{%
  \if@filesw
    \immediate\write\@auxout{\string\citation{apsrev41Control}}%
  \fi
}
\makeatother

\begin{document}

\title{Inelastic Scalar Dark Matter in a Gauged
   \texorpdfstring{$U(1)_{B-L}$}{U(1)(B-L)} Model and its LZ Phenomenology}

\author{Changjie Dai}
\email{changjiedai@mail.bnu.edu.cn}
\thanks{Changjie Dai, Linyun He, and Yonghua Wang share first authorship.}
\affiliation{Key Laboratory of Multi-scale Spin Physics, Ministry of Education,
  Beijing Normal University, Beijing 100875, China}
\affiliation{Center of Advanced Quantum Studies, School of Physics and Astronomy,
  Beijing Normal University, Beijing 100875, China}

\author{Linyun He}
\email{linyunhe@mail.bnu.edu.cn}
\affiliation{Key Laboratory of Multi-scale Spin Physics, Ministry of Education,
  Beijing Normal University, Beijing 100875, China}
\affiliation{Center of Advanced Quantum Studies, School of Physics and Astronomy,
  Beijing Normal University, Beijing 100875, China}

\author{Yonghua Wang}
\email[]{yonghuawang@mail.bnu.edu.cn}
\affiliation{Key Laboratory of Multi-scale Spin Physics, Ministry of Education,
  Beijing Normal University, Beijing 100875, China}
\affiliation{Center of Advanced Quantum Studies, School of Physics and Astronomy,
  Beijing Normal University, Beijing 100875, China}

\author{Wei Chao}
\email{chaowei@bnu.edu.cn}
\affiliation{Key Laboratory of Multi-scale Spin Physics, Ministry of Education,
  Beijing Normal University, Beijing 100875, China}
\affiliation{Center of Advanced Quantum Studies, School of Physics and Astronomy,
  Beijing Normal University, Beijing 100875, China}

\begin{abstract}
Motivated by the high-energy nuclear-recoil candidate event at $248 \pm 23\,\mathrm{(stat)} \pm 23\,\mathrm{(sys)}\,\mathrm{keV}$ recently reported by the LUX-ZEPLIN (LZ) experiment, we investigate inelastic scalar dark matter (DM) in a gauged $U(1)_{B-L}$ extension of the Standard Model. The dark sector consists of two complex scalar singlets, $\Phi$ and $S$, where spontaneous breaking of $U(1)_{B-L}$ splits $\Phi$ into two nearly degenerate real scalars $\phi_1$ and $\phi_2$ ($m_{\phi_1} > m_{\phi_2}$) with gauge interactions mediated by the $U(1)_{B-L}$ gauge boson $A'$. Depending on the lifetime of the heavier state $\phi_1$ for the decay $\phi_1 \to \phi_2 \nu \bar{\nu}$, this framework can realize either an exothermic ($\phi_1 N \to \phi_2 N$) or an endothermic ($\phi_2 N \to \phi_1 N$) scattering regime. We examine the thermal relic abundance of the dark scalars, calculate the predicted nuclear-recoil spectra in LZ under both the Standard Halo Model (SHM) and a Milky Way plus Large Magellanic Cloud (MW+LMC) halo model, and perform a profile-likelihood analysis incorporating the heavier-state lifetime or depletion conditions, the relic-density requirement, and the collider bound. Although the gauged $U(1)_{B-L}$ model accommodates both exothermic and endothermic DM, we find that the LZ data strongly favor the endothermic scenario, whereas the exothermic realization struggles to explain the observed signal. In the exothermic regime, requiring $\phi_1$ to survive until today severely suppresses the nuclear-recoil cross section, leaving fewer than $10^{-3}$ events in the $[215.5, 280.5]\,\mathrm{keV}$ signal interval. By contrast, the endothermic scenario is unsuppressed by the heavy-state depletion condition and naturally accounts for the LZ event for $m_\phi \sim 1.0$--$1.6\,\mathrm{TeV}$ and $\Delta m \sim 376$--$497\,\mathrm{keV}$, with the LMC high-velocity tail further extending the preferred parameter space toward larger mass splittings.
\end{abstract}

\maketitle
\clearpage

\section{Introduction}
Astrophysical and cosmological observations provide strong evidence for dark matter (DM), which accounts for about 26\% of the energy density of the Universe~\cite{Planck:2018vyg}; however, its particle nature remains unknown.
Weakly interacting massive particles (WIMPs) are well-motivated candidates, as their thermal freeze-out can naturally account for the observed DM relic abundance.
The possible interactions between WIMPs and Standard Model (SM) particles have motivated numerous direct-detection experiments, including LUX-ZEPLIN (LZ)~\cite{LZ:2022lsv,LZ:2024zvo}, XENONnT~\cite{XENON:2023cxc}, and PandaX-4T~\cite{PandaX-4T:2021bab}.
These results have imposed stringent constraints on the DM mass and its cross section with nuclei, motivating the study of interactions that produce recoil spectra different from those of conventional elastic WIMPs.

Recently, the LZ Collaboration has reported a high-energy nuclear-recoil candidate in an exposure of $2.84$~tonne-years~\cite{LZ:2026axp}. The reconstructed recoil energy is $248\pm23\,\mathrm{(stat)}\pm23\,\mathrm{(sys)}\,\mathrm{keV}$ with a global significance of $2.6\sigma$~\cite{LZ:2026axp}, in a region with a low expected background. The isolated high-energy event has motivated DM interpretations with recoil spectra that suppress the accompanying low-energy signal.
The conventional elastic, spin-independent DM model predicts a recoil energy spectrum that is concentrated at low recoil energies and therefore struggles to explain this high-energy excess.
Several DM interpretations have therefore been proposed, including momentum-dependent elastic DM~\cite{Unwin:2026rdp,Arcadi:2026kev,CentellesChulia:2026ats}, spin-dependent elastic DM~\cite{Elahi:2026vlm,Roy:2026wzb}, boosted DM~\cite{Liang:2026coz,Kannike:2026qyl,Heikinheimo:2026kwp,Chauhan:2026udz}, elastic DM with non-trivial form factors~\cite{Khan:2026osp,Jung:2026otm}, and inelastic DM~\cite{Tucker-Smith:2001myb,Tucker-Smith:2004mxa,Fan:2026hzw}.
Anwar et al.~\cite{Anwar:2026lzlhc} tested the elastic DM interpretation of the LZ event by recasting the ATLAS monojet search with missing transverse momentum. Within the validity of the effective field theory, they found that Large Hadron Collider (LHC) data disfavor the LZ-compatible region at DM masses below about $1\,\mathrm{TeV}$ for most of the interactions considered.

Specifically, realizations of inelastic DM have been studied in Higgsino scenarios~\cite{Fan:2026kxx,Freese:2026sga,Wu:2026nhi,Yin:2026jnn,DiMauro:2026ldr,Chattopadhyay:2026ryw,Du:2026guj,Bisal:2026khf,Langhoff:2026ujr,Frolovsky:2026tvq,Khan:2026zuj,Unwin:2026qep,Pallis:2026jdc,Khan:2026pgu,Hor:2026voa,Jeong:2026tlf,Lian:2026bze}, extended $U(1)$ gauge symmetries~\cite{Du:2026lpa,DiMauro:2026syh,Gemmell:2026yaw,Sahoo:2026dnw,Okada:2026upm,Das:2026buc,Pradhan:2026inq,Zhu:2026dag,Tang:2026lgu,Jodlowski:2026wjx,Lee:2026xxh,Ahmed:2026com,Sheikh:2026lxx,Okada:2026fef,Kumar:2026lgi,Okada:2026eol,Yamashita:2026ump,deLima:2026shq,Baer:2026fpy,Lee:2026wof,Yang:2026wpb,Qi:2026vyp,Barman:2026omh,He:2026idw,De:2026win,Sannino:2026hkc,Chanda:2026wdl,Chakraborti:2026gmu,Ahmed:2026jfv,Deka:2026zeu,Kim:2026euj}, and further inelastic-DM interpretations~\cite{Su:2026rwz,Nomura:2026qyq,Visinelli:2026kgt,Smirnov:2026aqk,Dent:2026bji,Wang:2026ytg,Das:2026uyy,Alhazmi:2026efz,Ahmed:2026qjg,Bandyopadhyay:2026gjw,Borah:2026zwf,Yuan:2026djt,Asadi:2026iot,Lee:2026jxl,He:2026hqz,Cabo-Almeida:2026uqw,Borah:2026ris,Lian:2026hpm,An:2026pkc,Xing:2026civ,Nagata:2026pbj,Paul:2026okh,Sheng:2026tqt,Girmohanta:2026qzo,Langhoff:2026qgo,Murayama:2026apt,Jho:2026cso,Jho:2026zzj,Choudhury:2026qjz,Lee:2026tro,Cerdeno:2026qio,Ibe:2026hyg}.
A complementary grand-unified realization based on $SU(6)$ correlates the inelastic mass splitting with a potentially collider-accessible $Z'$ gauge boson~\cite{Kotlarski:2026pep}.

In this inelastic scenario, DM transitions between two nearby states with a small mass splitting via scattering.
For endothermic scattering, the incoming particle makes a transition to a heavier state and must supply the excitation energy. This raises the minimum incident velocity and suppresses low-energy nuclear recoils, making the signal sensitive to the high-velocity tail of the halo distribution~\cite{Chang:2008gd,Bramante:2016rdh}.
For exothermic scattering, the incoming heavier state $\phi_1$ scatters off nuclei into the lighter state $\phi_2$, releasing the mass splitting $\Delta m$ as kinetic energy. At a characteristic nonzero recoil energy determined by the mass splitting and the DM--nucleus mass ratio, the minimum incident speed vanishes. Slow halo particles can therefore contribute to recoils near this energy, reducing the reliance on the high-velocity tail~\cite{Baer:2026fpy}. Recoils at sufficiently low energies remain kinematically suppressed~\cite{Graham:2010ca,Dent:2026bji,deLima:2026shq}.
%

An exothermic thermal-relic interpretation has also been developed with a complex scalar Higgs portal, connecting the LZ recoil to complementary photon signatures~\cite{Roy:2026bqn}.
In addition, solar capture and the resulting neutrino signals provide complementary tests of inelastic interpretations, including Higgsino-like DM~\cite{Bose:2026ndd,Nguyen:2026lui}, as well as elastic alternatives~\cite{Bose:2026szs}. Comparisons of fermionic and scalar inelastic candidates with matched xenon recoil responses further illustrate the model dependence of post-capture evolution and neutrino production~\cite{Capolupo:2026uyi}.

A gauged $U(1)_{B-L}$ extension of the SM offers a simple framework for scalar inelastic DM. Three right-handed neutrinos cancel the gauge anomalies and allow neutrinos to acquire mass via the seesaw mechanism~\cite{Minkowski:1977sc,Yanagida:1979as,Mohapatra:1979ia}. We consider two complex scalar singlets, $\Phi$ and $S$, with $B-L$ charges $Q_\Phi$ and $2Q_\Phi$, respectively. The vacuum expectation value of $S$ breaks $U(1)_{B-L}$, leaving a remnant parity that protects the lightest dark state~\cite{Krauss:1988zc,Petersen:2009ip,Batell:2010bp}. The interaction $S^\dagger\Phi^2+\mathrm{h.c.}$ splits $\Phi$ into two real scalars, $\phi_1$ and $\phi_2$, with $m_{\phi_1}>m_{\phi_2}$. The gauge boson $A'$ couples these states off-diagonally and mediates their inelastic scattering off SM particles. Related scalar-DM constructions based on broken gauge symmetries have been studied in Refs.~\cite{Baek:2014kna,Rodejohann:2015lca,Okada:2019sbb,Tang:2026lgu,Sheikh:2026lxx}.

Inelastic scalar DM in gauged $U(1)_{B-L}$ extensions has been explored as a possible interpretation of the high-energy recoil event reported by LZ. Ref.~\cite{Okada:2026eol} identified thermal-relic solutions near the gauge-boson resonance and found that the predicted DM--nucleon cross section can reach the scale suggested by the event, while also discussing complementary collider tests. Ref.~\cite{Kumar:2026lgi} studied the same inelastic mechanism in a generalized chiral $U(1)_{B-L}$ framework and identified a benchmark compatible with the observed relic abundance and collider constraints. These studies connect the proposed LZ interpretation with thermal DM production and collider phenomenology, motivating a quantitative assessment of the predicted recoil spectrum and event rate.

Building on these studies, we perform a systematic and quantitative investigation of inelastic scalar DM in the gauged $U(1)_{B-L}$ model in light of the LZ high-energy recoil event, comparing both the exothermic and endothermic scattering regimes within a unified framework. We first analyze the thermal evolution of the dark sector to map out the parameter space reproducing the observed relic abundance $\Omega_{\phi_1+\phi_2} h^2 = 0.12$ and identify the dominant annihilation and coannihilation channels. Crucially, because the heavier-state decay width $\Gamma_{\phi_1 \to \phi_2 \nu\bar{\nu}}$ and the heavy-mediator DM--nucleus scattering cross section share the same parametric dependence on $g_X g_\phi / m_{A'}^2$, the cosmological lifetime requirement for exothermic DM and the present-day depletion condition for endothermic DM have sharply opposite implications for direct detection. We then compute the predicted LZ nuclear-recoil spectra incorporating detector efficiency and energy resolution under both the Standard Halo Model (SHM) and a Milky Way plus Large Magellanic Cloud (MW+LMC) velocity distribution, and carry out a profile-likelihood analysis subject to the heavier-state lifetime or depletion constraints, the thermal relic-abundance requirement, and the LEP bound. Our results demonstrate that, while the gauged $U(1)_{B-L}$ model can accommodate both exothermic and endothermic DM, the LZ high-energy recoil event clearly favors the endothermic scenario, whereas the exothermic scenario is heavily suppressed by the lifetime constraint and can hardly account for the observed signal.

The rest of this paper is organized as follows. Sect.~\ref{sec:model} introduces the gauged $U(1)_{B-L}$ model and its interactions, and Sect.~\ref{sec:relic} examines the thermal evolution and relic abundance of the dark sector. In Sect.~\ref{sec:dd}, we calculate the inelastic nuclear-recoil spectra and describe the LZ detector response. Sect.~\ref{sec:like} presents the likelihood analysis under the heavy-state fraction, relic-abundance, and LEP constraints, and Sect.~\ref{sec:conclusion} summarizes our results.

\section{Dark matter model}\label{sec:model}
We extend the SM gauge group to $SU(3)_C \times SU(2)_L \times U(1)_Y \times U(1)_{B-L}$, and introduce three generations of right-handed neutrinos to cancel the gauge and gravitational anomalies~\cite{Basso:2008iv}.
In the dark sector, we introduce two complex scalar fields $\Phi$ and $S$, which are singlets under the SM gauge groups but carry charges under the $U(1)_{B-L}$ gauge group.
The $B-L$ charges of the particles are summarized in Table~\ref{tab:charges}.

\begin{table}[!htbp]
\caption{Charges of the particles under the gauge group $U(1)_{B-L}$. $Q_L$, $u_R$, $d_R$, $L_L$, $e_R$, and $N_R$ are the SM quarks and leptons and the right-handed neutrinos (three generations each), and $H$ is the SM Higgs doublet.}
\label{tab:charges}
\centering
\setlength{\tabcolsep}{4pt}
\begin{tabular}{@{}cccccccccc@{}}
\hline
 & $Q_L$ & $u_R$ & $d_R$ & $L_L$ & $e_R$ & $N_R$ & $H$ & $\Phi$ & $S$ \\
\hline
$Q_{B-L}$ & $1/3$ & $1/3$ & $1/3$ & $-1$ & $-1$ & $-1$ & $0$ & $Q_{\Phi}$ & $2 Q_{\Phi}$ \\
\hline
\end{tabular}
\end{table}
The relevant Lagrangian for $S$ and $\Phi$ is given by
\begin{align}
    \mathcal{L}_{S,\Phi} &= (D_\mu S)^\dagger (D^\mu S)
        + (D_\mu \Phi)^\dagger (D^\mu \Phi) \notag\\
    &\quad - V(S,\Phi),
\end{align}
where the covariant derivative is
$
    D_\mu = \partial_\mu - i g_X Q_{B-L} A'_\mu
$
with $Q_{B-L}(\Phi)=Q_\Phi$ and $Q_{B-L}(S)=2 Q_\Phi$. Here, 
 $g_{X}$ is the gauge coupling constant of $U(1)_{B-L}$, and $A'_\mu$ is the corresponding gauge field. The scalar potential $V(S,\Phi)$ is 
\begin{align}
    V(S,\Phi) &= -\mu_S^2|S|^2+\lambda_S|S|^4
        +m_\Phi^2|\Phi|^2+\lambda_\Phi|\Phi|^4 \notag\\
    &\quad+\lambda_{S\Phi}|S|^2|\Phi|^2
        +\frac{\kappa}{2}\left(S^\dagger\Phi^2+S\Phi^{\dagger 2}\right).
\end{align}
For this potential, the scalar $S$ develops a nonzero vacuum expectation value (VEV), $\langle S\rangle = v_S/\sqrt{2}$ with $v_S = \sqrt{\mu_S^2/\lambda_S}$, which spontaneously breaks the $U(1)_{B-L}$ gauge symmetry, whereas $\Phi$ does not acquire a VEV, $\langle \Phi\rangle = 0$, provided that $m_\Phi^2+\lambda_{S\Phi} v_S^2/2 > 0$.
After symmetry breaking, the two complex scalars can be parameterized as
\begin{align}
    S = \frac{v_S + s}{\sqrt{2}}, \quad
    \Phi = \frac{\phi_1 + i \phi_2}{\sqrt{2}},
\end{align}
where $s$, $\phi_1$, and $\phi_2$ are real scalar fields.
The Lagrangian becomes
\begin{align}
    \mathcal{L}_{S,\Phi} &= \mathcal{L}_{\mathrm{kin}} + \mathcal{L}_{\mathrm{mass}} + \mathcal{L}_{\mathrm{int}},
\end{align}
where $\mathcal{L}_{\mathrm{kin}}$ represents the kinetic terms, and $\mathcal{L}_{\mathrm{mass}}$ denotes the mass terms
\begin{align}
    \mathcal{L}_{\mathrm{mass}} &= \frac{1}{2}m_{A'}^2 A'_\mu A'^\mu -\frac{1}{2}m_s^2 s^2 \notag\\
    &\quad -\frac{1}{2}m_{\phi_1}^2 \phi_1^2-\frac{1}{2}m_{\phi_2}^2 \phi_2^2,
\end{align}
with
\begin{align}
    m_{A'} &= 2 g_\phi v_S, \qquad
    m_s^2 = 2\lambda_S v_S^2, \\
    m_{\phi_1}^2 &= m_\Phi^2+\frac{\lambda_{S\Phi}}{2}v_S^2+\frac{\kappa}{\sqrt2}v_S, \notag\\
    m_{\phi_2}^2 &= m_\Phi^2+\frac{\lambda_{S\Phi}}{2}v_S^2-\frac{\kappa}{\sqrt2}v_S,
\end{align}
where $g_\phi = g_X Q_\Phi$, and $\mathcal{L}_{\mathrm{int}}$ stands for the interaction terms
\begin{align}
    \mathcal{L}_{\mathrm{int}} &= -\lambda_S v_S\, s^3 -\frac{\lambda_S}{4}s^4
    -\frac{\lambda_\Phi}{4}(\phi_1^2+\phi_2^2)^2 \nonumber\\
    &\quad -\frac{\lambda_{S\Phi}}{2}v_S\, s(\phi_1^2+\phi_2^2) -\frac{\lambda_{S\Phi}}{4}s^2(\phi_1^2+\phi_2^2) \notag\\
    &\quad -\frac{\kappa}{2\sqrt2}s(\phi_1^2-\phi_2^2)
    +4 g_\phi^2 v_S\, s\, A'_\mu A'^\mu \notag\\
    &\quad +2 g_\phi^2 s^2 A'_\mu A'^\mu
    +\frac{1}{2}g_\phi^2 (\phi_1^2+\phi_2^2) A'_\mu A'^\mu \nonumber\\
    &\quad + g_\phi A'^\mu\left(\phi_2 \partial_\mu \phi_1 - \phi_1 \partial_\mu \phi_2\right). \label{Lint}
\end{align}
The two real scalars, $\phi_1$ and $\phi_2$, therefore have a mass splitting $m_{\phi_1}^2-m_{\phi_2}^2=\sqrt{2}\kappa v_S>0$ and an off-diagonal gauge interaction. Hereafter, we denote $m_\phi \equiv m_{\phi_2}$ and define $\Delta m\equiv m_{\phi_1}-m_{\phi_2}$.
Besides, the gauge boson $A'_\mu$ couples to the SM fermions and right-handed neutrinos through the $U(1)_{B-L}$ gauge interaction
\begin{equation}
    \mathcal{L}_{B-L} = g_X A'_\mu \sum_f Q_{B-L}^f \bar{f} \gamma^\mu f, \label{gauge_inter}
\end{equation}
where $Q_{B-L}^f$ is the $B-L$ charge of the fermion $f$, and the sum runs over all SM fermions and right-handed neutrinos.
Consequently, $\phi_1$ and $\phi_2$ can interact with a nucleus via $A'$ exchange.

\section{Dark matter relic abundance}\label{sec:relic}

In this section, we investigate the thermal evolution of dark sector particles in the early Universe, and identify the region in parameter space producing the correct DM abundance.

We assume that the dark sector particles are initially in thermal equilibrium with the SM particles. We study their evolution in the broken phase of $U(1)_{B-L}$, with the interactions given in Eq.~\eqref{Lint}.
We focus on the regime where $\kappa$ is small, leading to nearly degenerate dark scalars, $m_{\phi_1}\simeq m_{\phi_2}\equiv m_\phi$.
The particles $s$, $A'$, and $\phi_1$ are unstable, and they can decay through the following channels if kinematically allowed:
\begin{equation}
    \begin{gathered}
        s\to\phi_i\phi_i,\qquad s\to A'A',\qquad A'\to \phi_1\phi_2,\\
        A'\to f\bar f,\qquad \phi_1\to\phi_2\nu\bar\nu.
    \end{gathered}
\end{equation}
Here, we restrict $\Delta m < 2 m_e$, such that $\phi_1$ can only decay into $\phi_2$ plus neutrinos.
Whether DM scattering is exothermic or endothermic is governed by the lifetime of the heavier state, $\tau_1 = \Gamma_{\phi_1\to\phi_2\nu\bar\nu}^{-1}$, whose decay width is given by
\begin{align}
   \tau_1^{-1} = \Gamma_{\phi_1\to\phi_2\nu\bar\nu}
    &\simeq
    \frac{g_X^2 g_\phi^2 (\Delta m)^5}{40\pi^3m_{A'}^4},
\end{align}
where we have included all three active neutrino flavors.

Requiring the lifetime of $\phi_1$ to exceed the age of the Universe, $\tau_1 > t_0=13.8\, \text{Gyr}$, translates into an upper bound:
\begin{equation}
    \frac{g_X g_\phi}{m_{A'}^2} < 4.33 \times 10^{-10} \,\text{GeV}^{-2}\times\left(\frac{100\,\mathrm{keV}}{\Delta m}\right)^{5/2}.
    \label{eq:exoDMlifetime}
\end{equation}
In this case, both $\phi_1$ and $\phi_2$ contribute to the present-day DM abundance. 
Despite the small population of the heavy state $\phi_1$, exothermic DM--nucleus scattering can dominate the recoil spectrum owing to its lower minimum-velocity threshold compared to the corresponding endothermic process~\cite{Batell:2009vb,Finkbeiner:2009mi,Fan:2026hzw,Oliveira:2026jtn}.

Alternatively, if we require endothermic scattering to dominate, the present-day fraction of $\phi_1$ must be sufficiently suppressed.
This requires that the fraction of $\phi_1$ in the present-day total DM abundance be less than $10^{-7}$~\cite{Fan:2026hzw}.
The present-day number fraction of $\phi_1$ is given by
\begin{equation}
    f_1(t_0)
    \equiv
    \frac{n_{\phi_1}(t_0)}
    {n_{\phi_1}(t_0)+n_{\phi_2}(t_0)}
    \simeq \frac{1}{2}e^{-\Gamma_{\phi_1}t_0} < 10^{-7}.
\end{equation}
This depletion requirement gives the lower bound
\begin{equation}
    \frac{g_X g_\phi}{m_{A'}^2}\gtrsim 1.70\times10^{-9} \,\text{GeV}^{-2} \times\left(\frac{100\,\mathrm{keV}}{\Delta m}\right)^{5/2}.
    \label{eq:thermal-gX-depletion-bound}
\end{equation} 

The evolution of the total DM abundance, $n_\phi=n_{\phi_1}+n_{\phi_2}$, is governed by the Boltzmann equation
\begin{equation}
    \frac{dn_\phi}{dt}+3Hn_\phi = - \langle \sigma v\rangle_{\text{eff}}\left(n_\phi^2-(n_\phi^{\text{eq}})^2\right), \label{Beq}
\end{equation}
where $H$ is the Hubble parameter, and $n_\phi^{\text{eq}}$ is the number density in equilibrium. $ \langle \sigma v\rangle_{\text{eff}} = (\langle\sigma v\rangle_{11}+\langle \sigma v\rangle_{22}+2\langle \sigma v\rangle_{12})/4$, where $\langle \sigma v\rangle_{ij}$, $i,j=1,2$, is the thermally averaged annihilation/coannihilation cross section for $\phi_1$ and $\phi_2$.
The annihilation processes include $\phi_i\phi_i\leftrightarrow A'A'$ and
$\phi_i\phi_i\leftrightarrow ss$, and the coannihilation processes include $\phi_1\phi_2\leftrightarrow f\bar f$ and
$\phi_1\phi_2\leftrightarrow sA'$.
Since we are interested in the total DM abundance, we neglect the conversion processes $\phi_1\phi_1\leftrightarrow\phi_2\phi_2$ and $\phi_1 f\leftrightarrow\phi_2 f$, which maintain approximately equal abundances of the two dark scalars until the conversion freeze-out temperature $T_{\text{conv}}$, provided $T_{\text{conv}} \gg \Delta m$~\cite{Gemmell:2026yaw}.

\begin{figure}[!htbp]
    \centering
    \begin{minipage}[t]{0.49\linewidth}
        \centering
        \includegraphics[width=\linewidth]{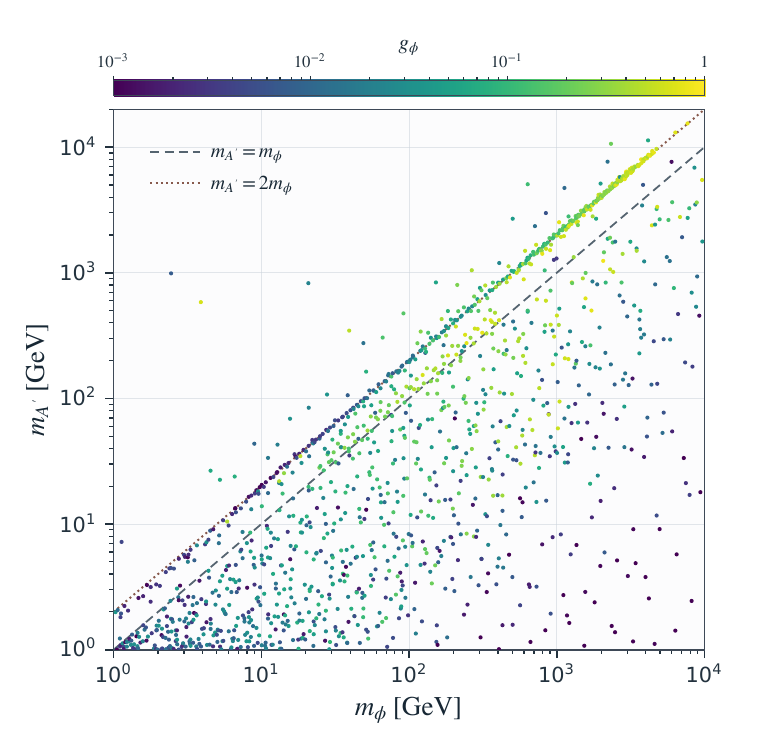}
    \end{minipage}\hfill
    \begin{minipage}[t]{0.49\linewidth}
        \centering
        \includegraphics[width=\linewidth]{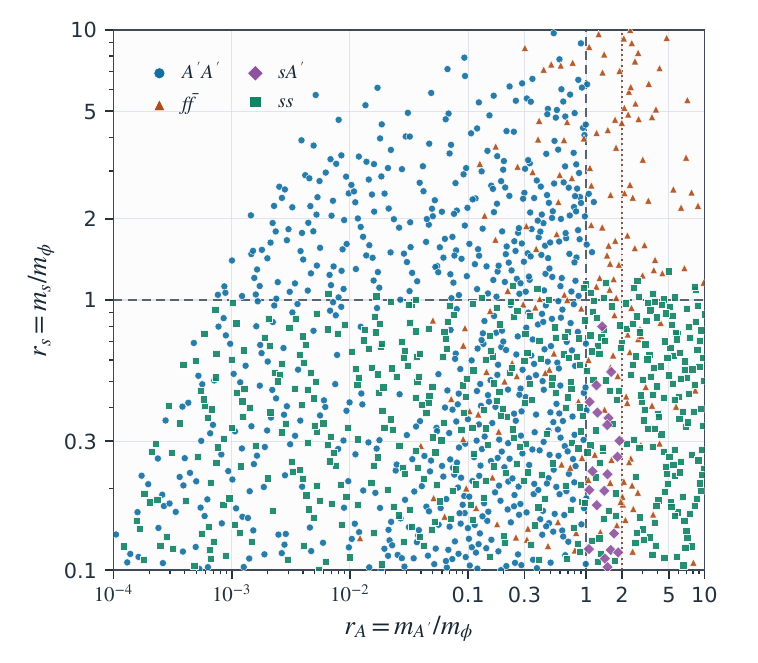}
    \end{minipage}
    \caption{Thermal relic-density solutions satisfying
    $\Omega_{\phi_1+\phi_2}h^2=0.12$. 
    \textbf{Left panel.}
    Projection onto the $(m_\phi,m_{A'})$ plane. The color gives the smallest $g_\phi$ found for each sampled mass pair. The dashed line at $m_{A'}=m_\phi$ marks the zero-velocity threshold for the $\phi_i \phi_i \to A'A'$ process, and the dotted line at $m_{A'}=2m_\phi$ marks the $A'$ resonance enhancement for the $\phi_1 \phi_2 \to f \bar f$ process.
    \textbf{Right panel.}
    Leading depletion channel in the mass-ratio plane
    $(r_A,r_s)=(m_{A'}/m_\phi,m_s/m_\phi)$. Each marker shows the final state with the largest contribution to the effective annihilation rate at freeze-out. The dashed lines at $r_A=1$ and $r_s=1$ mark the zero-velocity thresholds for DM annihilations with  $A'A'$ and $ss$ final states, respectively. The dotted line at $r_A=2$ marks the $A'$ resonance enhancement for coannihilation with $f \bar f$ final states. The $f\bar f$ channel is concentrated near the enhancement line, and the $sA'$ solutions are rare.}
    \label{fig:relic-plane}
\end{figure}

The left panel of Fig.~\ref{fig:relic-plane} shows the solutions that reproduce $\Omega_{\phi_1+\phi_2}h^2=0.12$ in the
$(m_\phi,m_{A'})$ plane. At each sampled mass pair, we search for combinations of $m_s$, $g_\phi$, and $\lambda_{S\Phi}$ that satisfy the relic-abundance condition. The color gives the smallest $g_\phi$ found among the solutions at that mass pair. The narrow band near $m_{A'}=2m_\phi$ arises mainly from resonant
coannihilation, $\phi_1\phi_2\to A'^\ast\to f\bar f$. Most of the sampled solutions lie below $m_{A'}=m_\phi$, where $\phi_i\phi_i\to A'A'$ is open at zero relative velocity. 
The right panel projects the relic-density solutions onto the $(r_A,r_s)=(m_{A'}/m_\phi,m_s/m_\phi)$ plane and identifies the channel with the largest contribution to the effective annihilation rate at freeze-out. The green squares, which represent $\phi_i\phi_i\to ss$, lie mainly at $r_s<1$, where the $ss$ final state is kinematically accessible at zero relative velocity. Similarly, the blue circles for $\phi_i\phi_i\to A'A'$ lie mostly at $r_A<1$. Among the sampled solutions, annihilations generally give the largest contribution where these channels are kinematically unsuppressed. Coannihilations are concentrated in more localized regions, particularly near $r_A=2$ and the $sA'$ production threshold.

\section{Direct detection}\label{sec:dd}
In this section, we investigate potential nuclear-recoil signals in the LZ detector and calculate the predicted nuclear-recoil event spectra for the exothermic and endothermic DM scenarios.

The inelastic scattering event rate in the recoil energy interval $[E_R, E_R + dE_R]$ per unit detector mass is given by
\begin{equation}
\frac{dR_T}{dE_R}=\frac{\rho_{\phi_i}}{m_Tm_\phi}\int_{|\mathbf{v}|\geq v_{\mathrm{min}}(E_R)}d^3v\,f_{\mathrm{lab}}(\mathbf{v})\,v\,\frac{d\sigma_T}{dE_R}.
\label{eq:true-recoil-rate}
\end{equation}
Here, $E_R$ is the nuclear recoil energy, $m_T$ is the target nucleus mass, and $\rho_{\phi_i}$ is the density of the incoming DM component $\phi_i$ ($i=1,2$). The component density is given by $\rho_{\phi_i}=f_i \rho_{\mathrm{DM}}$, where $f_i$ is the fraction of the total local DM density contributed by component $i$. We adopt a benchmark value of $\rho_{\mathrm{DM}}=0.3\,\mathrm{GeV/cm^3}$ for the total local DM density~\cite{Lewin:1995rx}.
To assess the dependence on the local DM velocity distribution, we consider two astrophysical cases: the Standard Halo Model (SHM) and a local DM distribution in the Milky Way (MW) that includes the effects of the Large Magellanic Cloud (LMC), hereafter denoted by MW+LMC. These two halo cases are considered separately from the endothermic and exothermic scattering scenarios. The normalized laboratory-frame velocity distribution is $f_{\mathrm{lab}}(\mathbf{v})=f_{\mathrm{gal}}(\mathbf{v}+\mathbf{v}_E)$, where $\mathbf{v}_E$ is the detector velocity relative to the Galactic frame. In the SHM, the Galactic-frame distribution is a truncated Maxwellian~\cite{Lewin:1995rx,McCabe:2010zh}:
\begin{equation}
    f_{\mathrm{gal}}(\mathbf{v}) =
    \begin{cases}
        \dfrac{1}{N_{\mathrm{esc}}} \left(\dfrac{1}{\pi v_0^2}\right)^{3/2} e^{-\mathbf{v}^2/v_0^2}, & |\mathbf{v}| < v_{\mathrm{esc}}, \\[6pt]
        0, & \text{otherwise.}
    \end{cases}
\end{equation}
For the SHM benchmark, we use $v_0=238\,\mathrm{km/s}$, $v_{\mathrm{esc}}=544\,\mathrm{km/s}$, and $|\mathbf{v}_E|=253.7\,\mathrm{km/s}$, without annual modulation. These fixed halo inputs have uncertainties~\cite{McCabe:2010zh,Piffl:2013mla}, which can be important for endothermic scattering near its velocity threshold~\cite{Fan:2026kxx}. The normalization factor is~\cite{Freese:2012xd}
\begin{equation}
    N_{\mathrm{esc}} = \mathrm{erf}(z) - \frac{2}{\sqrt{\pi}} z e^{-z^2},
\end{equation}
where $z\equiv v_{\mathrm{esc}}/v_0$.

In the MW+LMC case, we instead consider a local velocity distribution that includes both DM originating from the LMC and the dynamical response of the Milky Way halo to its infall. Cosmological simulations show that these effects can enhance the high-speed tail in the Solar neighborhood: LMC particles can populate the fastest part of the distribution, while native Milky Way particles can also be accelerated~\cite{SmithOrlik:2023LMC}. Thus, the LMC contribution should not be interpreted simply as an additional Maxwellian component superposed on an otherwise unchanged SHM. For the comparison, we retain the same benchmark total local DM density and encode the halo dependence in the normalized distribution $f_{\mathrm{lab}}^{\mathrm{MW+LMC}}$ entering Eq.~\eqref{eq:true-recoil-rate}. An enhanced high-speed population is particularly relevant for endothermic scattering, for which the incoming DM must supply the excitation energy; it can increase the rate at large minimum speeds and extend the kinematically accessible mass splitting~\cite{ReynosoCordova:2024LMC}. For exothermic scattering, the sensitivity to this tail depends on the recoil energy through $v_{\mathrm{min}}(E_R)$ and is reduced near the recoil energy at which $v_{\mathrm{min}}$ vanishes.
The enhanced high-speed tail is sensitive to the adopted halo reconstruction and LMC analog. Comparisons of the SHM, Au-25, and a Gaia-based Milky Way reconstruction combined with a FIRE-2 LMC analog show appreciable variation in direct-detection predictions~\cite{Zhang:2026hmg}; our MW+LMC results should therefore be interpreted as benchmark-dependent.
Here, $v_\mathrm{min}(E_R)$ is the minimum velocity required for a recoil energy $E_R$, and is given by~\cite{Tucker-Smith:2001myb,Graham:2010ca}
\begin{equation}
    v_\mathrm{min}(E_R) = \frac{1}{\sqrt{2 m_T E_R}} \left| \frac{m_T E_R}{\mu_T} \pm \Delta m \right|,
\end{equation}
where $\mu_T=m_\phi m_T/(m_\phi+m_T)$ is the reduced mass and the plus (minus) sign describes endothermic (exothermic) scattering.

\begin{figure}[!htbp]
    \centering
    \includegraphics[width=0.92\linewidth]{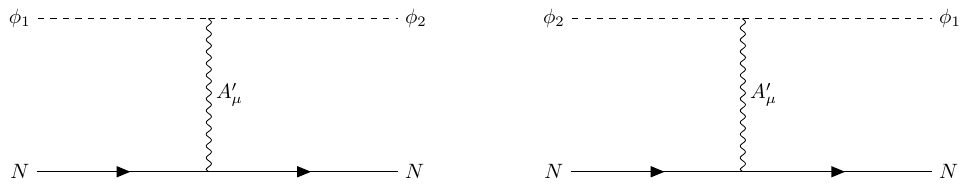}
    \caption{Feynman diagrams contributing to the inelastic DM--nucleus scattering cross section, mediated by $t$-channel exchange of the $U(1)_{B-L}$ gauge boson $A'_\mu$: (left) the exothermic process $\phi_1 N \to \phi_2 N$; (right) the endothermic process $\phi_2 N \to \phi_1 N$.}
    \label{fig:feynman_diagrams}
\end{figure}

The tree-level processes are shown in Fig.~\ref{fig:feynman_diagrams}. The differential cross section is
\begin{equation}
    \frac{d\sigma_T}{dE_R} = \frac{g_X^2 g_\phi^2 A^2 m_T}{2\pi v^2\left(m_{A'}^2+2m_T E_R\right)^2}F^2(E_R),
\end{equation}
where $A$ is the mass number of the target nucleus, and $F(E_R)$ is the Helm form factor~\cite{Helm:1956zz,Lewin:1995rx,Duda:2006uk}
\begin{align}
 F^2(E_R)&=\left[\frac{3j_1(q_Rr_N)}{q_Rr_N}\right]^2e^{-(q_Rs_N)^2},\\
 r_N&=\sqrt{c_N^2+\frac{7\pi^2a_N^2}{3}-5s_N^2}, \notag\\
 q_R&=\sqrt{2m_TE_R},\\
 c_N&=(1.23A^{1/3}-0.60)\,\mathrm{fm}, \notag\\
 a_N&=0.52\,\mathrm{fm},\qquad s_N=0.90\,\mathrm{fm}.
\end{align}
Here $j_1$ is the spherical Bessel function. We use $q_R$ for the momentum transfer to distinguish it from the likelihood-ratio statistic introduced below. In the heavy-mediator limit, the differential cross section is proportional to $g_X^2 g_\phi^2/m_{A'}^4$. Therefore, the cross section for exothermic DM scattering off nuclei is strongly suppressed by the lifetime constraint in Eq.~\eqref{eq:exoDMlifetime}, whereas that for endothermic scattering is not suppressed by the heavier-state depletion condition in Eq.~\eqref{eq:thermal-gX-depletion-bound}.

\begin{figure}[!htbp]
    \centering
    \begin{minipage}[t]{0.49\linewidth}
        \centering
        \includegraphics[width=\linewidth]{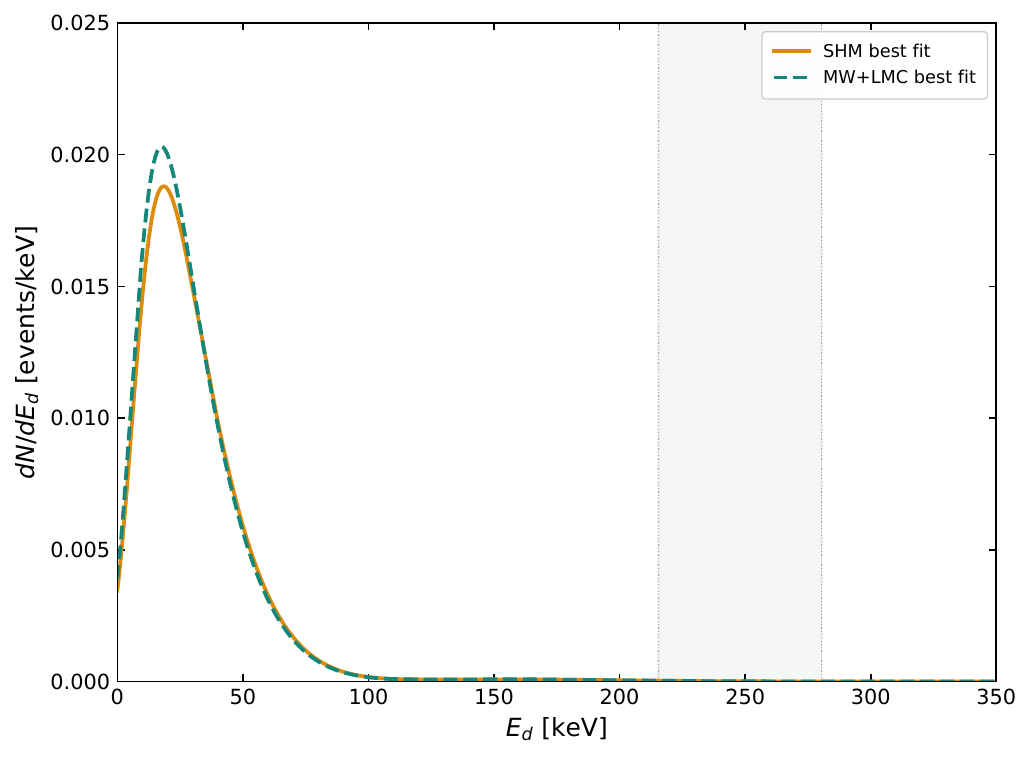}
    \end{minipage}\hfill
    \begin{minipage}[t]{0.49\linewidth}
        \centering
        \includegraphics[width=\linewidth]{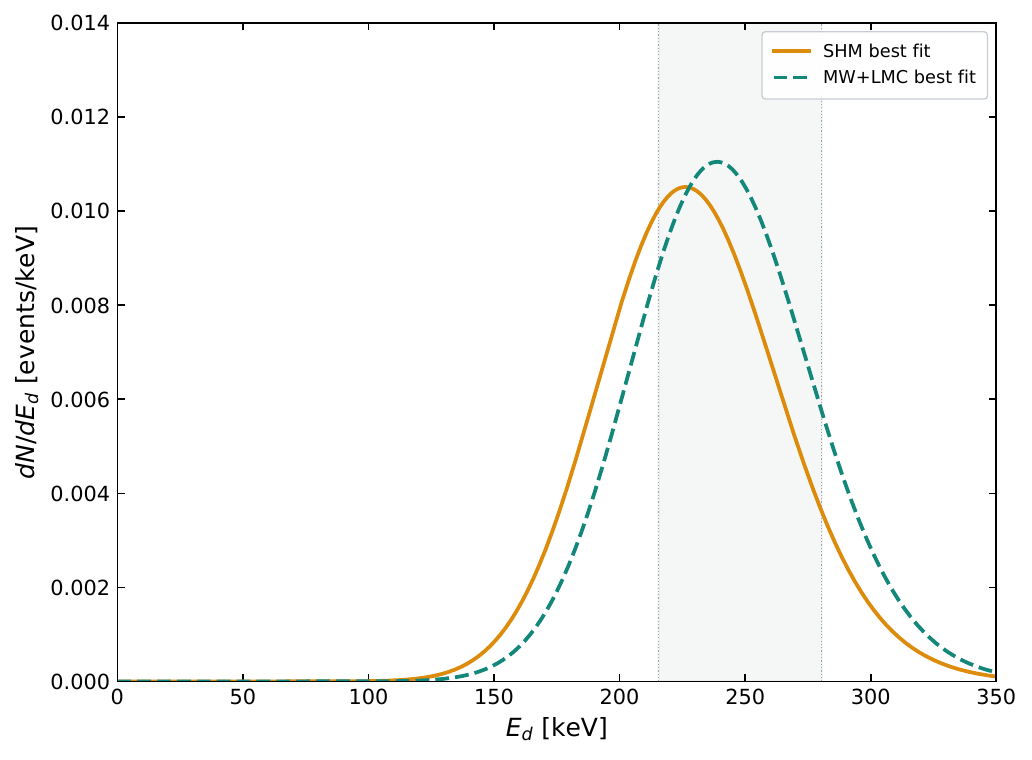}
    \end{minipage}
    \caption{Predicted recoil event spectra for exothermic (left) and endothermic (right) scattering in LZ, with the gray shaded region indicating the signal interval $[215.5, 280.5]\,\mathrm{keV}$. The dashed (labeled ``MW+LMC best fit'' in the legend) and solid curves correspond to the MW+LMC and SHM halo models, respectively.  In both scenarios, we choose $g_\phi=0.3$ to reproduce the observed DM abundance. The benchmark points shown below are the best-fit points obtained from the likelihood analysis in Sect.~\ref{sec:like}.
    \textbf{{Left panel:}} Exothermic DM scenario. For the SHM, we choose a benchmark point with $g_X=10^{-3}$, $m_{A'}=606.5\,\mathrm{GeV}$, $m_\phi=609.6\,\mathrm{GeV}$, and $\Delta m=76.3\,\mathrm{keV}$, which predicts $0.736$ total events, but only $6.3 \times 10^{-4}$ events fall in the signal interval. For the MW+LMC halo model, we choose a benchmark point with $g_X=3.5\times10^{-3}$, $m_{A'}=1129.1\,\mathrm{GeV}$, $m_\phi=533.4\,\mathrm{GeV}$, and $\Delta m=77.6\,\mathrm{keV}$, which predicts $0.764$ total events, but only $6.7 \times 10^{-4}$ events fall in the signal interval.
   \textbf{{Right panel:}} Endothermic DM scenario. For the SHM, we choose a benchmark point with $g_X=10^{-3}$, $m_{A'}=7\,\mathrm{GeV}$, $m_\phi=1.6\,\mathrm{TeV}$, and $\Delta m=376.4\,\mathrm{keV}$, which predicts $0.946$ total events, and $0.526$ events fall in the signal interval. For the MW+LMC halo model, we choose a benchmark point with $g_X=10^{-3}$, $m_{A'}=7\,\mathrm{GeV}$, $m_\phi=1\,\mathrm{TeV}$, and $\Delta m=496.5\,\mathrm{keV}$, which predicts $0.985$ total events, and $0.615$ events fall in the signal interval.}
    \label{fig:spectra}
\end{figure}

LZ has described its detector-response simulations and calibration systems in Refs.~\cite{LZ:2020zog,LZ:2024bsz}. More recent photoneutron measurements probe the nuclear-recoil response at low energies~\cite{LZ:2025vdo}.
We use the exposure $\mathcal E=2.84$~tonne-years from the LZ extended-energy search~\cite{LZ:2026axp}. The rate for natural xenon is $dR/dE_R=\sum_T\xi_T\,dR_T/dE_R$, where $\xi_T$ is the isotope mass fraction. Near the inelastic kinematic endpoint, differences in nuclear mass can make the isotope contributions substantially different~\cite{Ahmed:2026kan}. The expected spectrum in reconstructed recoil energy $E_d$ is
\begin{equation}
 \frac{dN}{dE_d}=\mathcal E\int_0^\infty dE_R\,
 \epsilon(E_R)\,G(E_d,E_R)\,\frac{dR}{dE_R}.
 \label{eq:response-convolution}
\end{equation}
Here $G$ is an approximate Gaussian response,
\begin{align}
 G(E_d,E_R)&=\frac{1}{\sqrt{2\pi}\sigma_E(E_R)}
 \exp\!\left[-\frac{(E_d-E_R)^2}{2\sigma_E^2(E_R)}\right],
\end{align}
where the energy resolution is given by~\cite{Fan:2026hzw}
\begin{align}
\sigma_E(E_R)=\sqrt{23^2+23^2}\sqrt{\frac{E_R}{248\,\mathrm{keV}}}\,\mathrm{keV}~.
\end{align}
Here $\epsilon(E_R)$ is the detector efficiency, which is digitized from Fig.~S2 of Ref.~\cite{LZ:2026axp}.

Fig.~\ref{fig:spectra} compares the predicted nuclear recoil spectra in LZ for the exothermic (left) and endothermic (right) scenarios at the best-fit benchmark points obtained from the likelihood analysis in Sect.~\ref{sec:like}, under both the SHM (solid) and MW+LMC (dashed) halo models. In the exothermic case, the strong suppression from the lifetime constraint leaves only $\mathcal{O}(10^{-4})$ events inside the $[215.5, 280.5]\,\mathrm{keV}$ signal interval (gray shaded region), despite predicting $\sim 0.75$ total events. Conversely, the endothermic scenario naturally populates the signal window, with $0.526$ ($0.615$) out of $0.946$ ($0.985$) total events falling within the signal interval for the SHM (MW+LMC) halo model. The high-velocity tail of the MW+LMC distribution further accommodates a larger mass splitting $\Delta m$ and enhances the signal-region yield compared to the SHM.

\section{Likelihood analysis of the LZ signal}\label{sec:like}

In this section, we perform a statistical hypothesis test of the DM model in both the exothermic and endothermic scenarios under the SHM and MW+LMC halo models, using the extended likelihood method~\cite{Barlow:1990vc,Fan:2026kxx}. We identify the region of parameter space that can explain the LZ experimental results while simultaneously reproducing the observed DM relic abundance.

We divide the reconstructed recoil energy $E_d$ into three intervals:
\begin{equation}
 \begin{aligned}
   \mathcal B_1&=[0,215.5]\,\mathrm{keV},\\
   \mathcal B_2&=[215.5,280.5]\,\mathrm{keV},\\
   \mathcal B_3&=[280.5,350]\,\mathrm{keV}.
 \end{aligned}
 \label{eq:lz-energy-bins}
\end{equation}
The middle bin contains the candidate event at $248\,\mathrm{keV}$~\cite{LZ:2026axp}. The observed bin counts are $(n_1,n_2,n_3)=(0,1,0)$. We integrate the predicted spectrum, including detector efficiency and energy resolution, over each bin and use only these bin counts in the likelihood (for related LZ analyses, see Refs.~\cite{Dent:2026bji,Palmisano:2026kuj}). Following Ref.~\cite{Fan:2026hzw}, we neglect the small expected background and write
\begin{equation}
	\begin{aligned}
	  \mathcal L_s(\boldsymbol{\theta})&=\prod_{a=1}^3
	    \frac{s_a^{n_a}e^{-s_a}}{n_a!}=s_2e^{-s_{\mathrm{tot}}},\\
	  s_{\mathrm{tot}}&=\sum_{a=1}^3 s_a,
	\end{aligned}
	\label{eq:extended-likelihood}
\end{equation}
where $s_a=\int_{\mathcal{B}_a}dE_d\,dN/dE_d$ is the expected signal count in bin $a$, with $dN/dE_d$ denoting the spectrum after detector response. The parameter vector $\boldsymbol{\theta}$ encompasses all adjustable parameters of the model, which in our case are $m_\phi$, $\Delta m$, $g_X$, and $m_{A'}$.
The DM mass and mass splitting primarily determine the shape of the recoil energy spectrum, whereas the other two parameters mainly affect the overall event rate. Therefore, we choose $m_\phi$ and $\Delta m$ as the parameters of interest in the hypothesis test, while treating the remaining two parameters, $g_X$ and $m_{A'}$, as nuisance parameters to be profiled over. In the standard profile likelihood approach, for each fixed pair of $m_\phi$ and $\Delta m$, we determine the values of $g_X$ and $m_{A'}$ that maximize the likelihood function. Scanning over the entire $(m_\phi,\Delta m)$ parameter space then yields the profiled values $g_X(m_\phi,\Delta m)$ and $m_{A'}(m_\phi,\Delta m)$ that maximize the likelihood at each point. We fix $g_\phi=0.3$ in the profiling procedure such that the observed relic abundance is reproduced.

Furthermore, in our profiling procedure, we maximize the likelihood only within the parameter region allowed by the constraints applied to each scenario. We take into account the constraints on the lifetime of the heavier state in Eq.~\eqref{eq:exoDMlifetime} and its depletion in Eq.~\eqref{eq:thermal-gX-depletion-bound} for the exothermic and endothermic scenarios, respectively, as well as the thermal relic abundance requirement for DM and the LEP bound ($m_{A'}/g_X\geq 7\,\mathrm{TeV}$)~\cite{Cacciapaglia:2006pk,Basso:2009hf}.
Finally, we obtain the profiled likelihood function $\mathcal L_{\mathrm{prof}}(m_\phi,\Delta m)$.

The statistic is
\begin{equation}
 \begin{aligned}
   q(m_\phi,\Delta m)&=-2\ln
     \frac{\mathcal L_{\mathrm{prof}}(m_\phi,\Delta m)}{\mathcal L_{\max}},\\
   \mathcal L_{\max}&=\max_{m_\phi,\Delta m}\mathcal L_{\mathrm{prof}}.
 \end{aligned}
\end{equation}
Each experimental realization yields a different dataset and, consequently, a different likelihood function. However, according to Wilks' theorem~\cite{Wilks:1938dza,Cowan:2010js}, the likelihood-ratio test statistic \(q(m_\phi,\Delta m)\) asymptotically follows a $\chi^2$ distribution under certain regularity conditions. In other words, for a given pair of DM mass and mass splitting, the $\chi^2$ distribution describes the probability density of the test statistic \(q(m_\phi,\Delta m)\) obtained from repeated experimental realizations, assuming that the corresponding model parameters are the true values.
If the observed value of \(q\) obtained from the LZ experimental data lies in the upper tail of the distribution predicted by a particular parameter point, this indicates that the corresponding model parameters are disfavored at a given confidence level, as they do not provide a sufficiently good description of the observed data. Therefore, at the \(1\sigma\) confidence level, parameter points satisfying \(q>2.30\) are excluded, whereas those satisfying \(q\le2.30\) remain allowed.

\begin{figure}[!htbp]
	\centering
    \begin{minipage}[t]{0.49\linewidth}
        \centering
        \includegraphics[width=\linewidth]{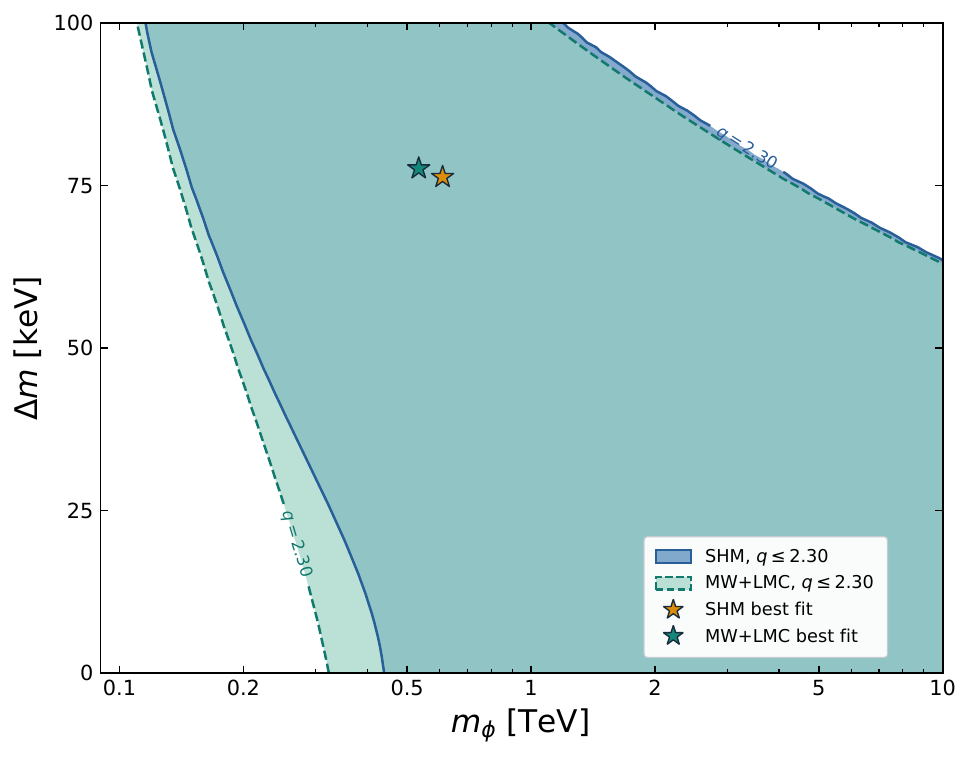}
    \end{minipage}\hfill
    \begin{minipage}[t]{0.49\linewidth}
        \centering
        \includegraphics[width=\linewidth]{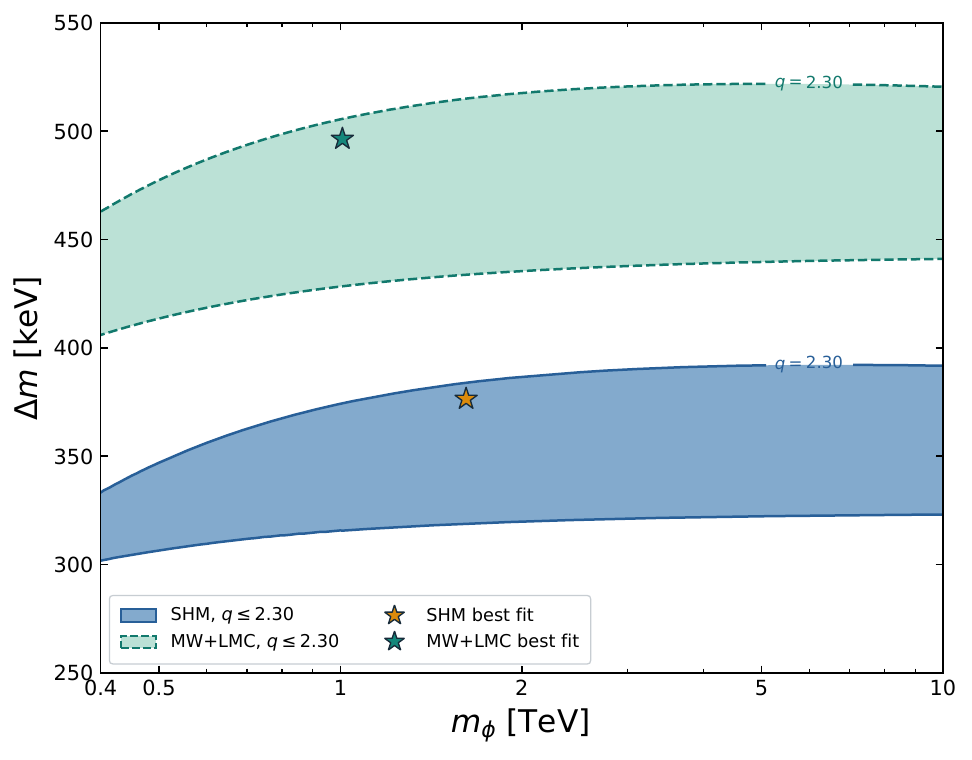}
    \end{minipage}
	\caption{Profile-likelihood contours in the $(m_\phi, \Delta m)$ plane after profiling over $m_{A'}$ and $g_X$, subject to the constraints on the $\phi_1$ lifetime and its present-day fraction for the exothermic (left) and endothermic (right) scenarios, respectively, together with the LEP bound and the DM relic-abundance requirement. The blue and green regions correspond to $q\le2.30$ under the SHM and the MW+LMC halo model, respectively, while the gold and green stars mark the corresponding best-fit points.}
	\label{fig:joint-confidence}
\end{figure}

Thus, we can determine the allowed region of the model parameter space through the hypothesis test.
Fig.~\ref{fig:joint-confidence} presents the resulting profile-likelihood contours in the $(m_\phi, \Delta m)$ plane for the exothermic (left panel) and endothermic (right panel) scenarios, obtained after profiling over $m_{A'}$ and $g_X$ subject to the respective heavier-state lifetime or depletion constraints, the LEP bound, and the thermal relic-abundance requirement (with $g_\phi = 0.3$). The blue and green shaded regions delineate the $1\sigma$ preferred parameter space ($q < 2.30$) under the SHM and MW+LMC halo models, respectively, while the gold and green stars indicate the corresponding best-fit points, whose parameters and predicted LZ event yields are summarized in Table~\ref{tab:benchmark_points}.

\begin{table}[!htbp]
\caption{Best-fit benchmark parameters and predicted LZ recoil event yields for exothermic and endothermic DM under the SHM and MW+LMC halo models, with $g_\phi = 0.3$ fixed to reproduce the observed relic abundance. Here $s_{\text{tot}}$ denotes the total number of predicted events and $s_{\text{sig}}$ denotes the number of events falling within the signal interval $[215.5, 280.5]\,\mathrm{keV}$.}
\label{tab:benchmark_points}
\centering
\small
\setlength{\tabcolsep}{5pt}
\begin{tabular}{@{}llrrrrrr@{}}
\hline
Scenario & Halo
& \shortstack{$m_\phi$\\{}[GeV]}
& \shortstack{$\Delta m$\\{}[keV]}
& \shortstack{$m_{A'}$\\{}[GeV]}
& $g_X$
& $s_{\text{tot}}$
& $s_{\text{sig}}$ \\
\hline
Exothermic & SHM    & 609.6 & 76.3  & 606.5 & $1.0\times 10^{-3}$ & 0.736 & $6.3\times 10^{-4}$ \\
           & MW+LMC & 533.4 & 77.6  & 1129.1    & $3.5\times 10^{-3}$ & 0.764 & $6.7\times 10^{-4}$ \\[3pt]
Endothermic & SHM    & 1616.8  & 376.4 & 7.0    & $1.0\times 10^{-3}$ & 0.946 & 0.526 \\
            & MW+LMC & 1007.7  & 496.5 & 7.0    & $1.0\times 10^{-3}$ & 0.985 & 0.615 \\
\hline
\end{tabular}
\end{table}

A clear contrast emerges between the two scenarios. In the exothermic case, the lifetime constraint severely restricts the viable parameter space and suppresses the signal rate in the high-energy interval, pushing the best-fit points to moderate DM masses ($m_\phi \sim 530$--$610\,\mathrm{GeV}$) and small mass splittings ($\Delta m \sim 76$--$78\,\mathrm{keV}$), where fewer than $10^{-3}$ events fall inside the $[215.5, 280.5]\,\mathrm{keV}$ window. In contrast, the endothermic scenario is not suppressed by the heavier-state depletion condition and favors heavier DM masses ($m_\phi \sim 1.0$--$1.6\,\mathrm{TeV}$) together with substantially larger mass splittings ($\Delta m \sim 376$--$497\,\mathrm{keV}$), yielding $s_{\text{sig}} = 0.526$ ($0.615$) events in the signal interval for the SHM (MW+LMC) halo model. Moreover, including the LMC high-velocity component shifts the preferred endothermic region toward larger mass splittings, as faster DM particles can more readily overcome the kinematic threshold for endothermic upscattering.

\section{Conclusions and Remarks}
\label{sec:conclusion}
In this paper, we have investigated inelastic scalar dark matter in a minimal gauged $U(1)_{B-L}$ extension of the Standard Model as a possible explanation of the high-energy nuclear-recoil candidate event at $248\,\mathrm{keV}$ reported by LZ. Introducing two complex scalar singlets $\Phi$ and $S$ with $B-L$ charges $Q_\Phi$ and $2Q_\Phi$, spontaneous breaking of $U(1)_{B-L}$ leaves a remnant symmetry stabilizing the lighter real scalar $\phi_2$ and generates a small mass splitting $\Delta m = m_{\phi_1} - m_{\phi_2}$ together with an off-diagonal coupling to the gauge boson $A'$. Focusing on the regime $\Delta m < 2m_e$, where the heavier state decays exclusively via $\phi_1 \to \phi_2 \nu \bar{\nu}$, we studied the early-Universe thermal evolution and identified the parameter space satisfying $\Omega_{\phi_1+\phi_2} h^2 = 0.12$. The relic abundance is primarily controlled by the annihilation channels $\phi_i\phi_i \to A'A'$ and $\phi_i\phi_i \to ss$ whenever kinematically accessible, complemented by resonant coannihilation $\phi_1\phi_2 \to A'^* \to f\bar{f}$ near $m_{A'} \simeq 2m_\phi$.

A central feature of this $U(1)_{B-L}$ framework is that it can accommodate both exothermic and endothermic DM depending on the lifetime of $\phi_1$, yet the LZ direct-detection data strongly distinguish between the two regimes. In the heavy-mediator limit, both the decay width $\Gamma_{\phi_1 \to \phi_2 \nu\bar{\nu}}$ and the inelastic DM--nucleus scattering cross section scale as $g_X^2 g_\phi^2 / m_{A'}^4$. For the exothermic scenario, requiring $\tau_1 > t_0$ so that $\phi_1$ survives until the present day places a stringent upper bound on $g_X g_\phi / m_{A'}^2$, which severely suppresses the scattering rate off xenon nuclei. Consequently, our profile-likelihood analysis, in which $m_{A'}$ and
$g_X$ are profiled over subject to the lifetime, relic-density,
and LEP ($m_{A'}/g_X \geq 7\,\mathrm{TeV}$) constraints, shows
that the exothermic best-fit points ($m_\phi \sim
530$--$610\,\mathrm{GeV}$, $\Delta m \sim 76$--$78\,\mathrm{keV}$)
predict fewer than $10^{-3}$ events inside the
$[215.5, 280.5]\,\mathrm{keV}$ signal interval for both halo models,
rendering the exothermic realization in this model unable to
account for the LZ event.

In contrast, the endothermic scenario requires the heavier state to be depleted today ($f_1(t_0) < 10^{-7}$), which imposes a lower bound on $g_X g_\phi / m_{A'}^2$ and therefore does not suppress the nuclear-recoil cross section. Within the constraints considered, endothermic scattering naturally produces a recoil spectrum peaking across the LZ signal window. At the $1\sigma$ preferred region ($q < 2.30$), the endothermic scenario favors heavier DM masses ($m_\phi \sim 1.0$--$1.6\,\mathrm{TeV}$) and larger mass splittings ($\Delta m \sim 376$--$497\,\mathrm{keV}$), yielding $s_{\text{sig}} = 0.526$ ($0.615$) events in the signal interval out of $0.946$ ($0.985$) total events under the SHM (MW+LMC) halo model. Furthermore, incorporating the high-velocity tail from the LMC shifts the preferred endothermic parameter space toward larger $\Delta m$ and enhances the signal-region yield, although this astrophysical enhancement remains sensitive to the adopted halo reconstruction.

Finally, it is worth noting that a similar inelastic scalar DM construction can also be realized under a gauged $U(1)_R$ symmetry \cite{Chao:2017rwv,Nomura:2017tih}, in which only right-handed fermions carry $U(1)_R$ charges. Because left-handed active neutrinos are neutral under $U(1)_R$ while right-handed quarks remain charged, the tree-level decay $\phi_1 \to \phi_2 \nu\bar{\nu}$ is absent and the mixing effects can be avoided by dimension-five Yukawa operators\cite{Dutta:2020enk}. This decouples the lifetime of the heavier state $\phi_1$ from the inelastic DM--nucleus scattering cross section, relaxing the stringent suppression encountered in the exothermic $U(1)_{B-L}$ regime. Consequently, a gauged $U(1)_R$ extension can simultaneously accommodate both exothermic and endothermic interpretations of the LZ high-energy recoil event. A dedicated investigation of the $U(1)_R$ realization, together with upcoming high-exposure data from LZ, XENONnT, and PandaX-4T as well as complementary collider searches, will further clarify the viable inelastic DM explanations of the LZ signal.

\section*{Statements and Declarations}

\subsection*{Funding}
This work was supported in part by the National Key R\&D Program of China under Grant No. 2023YFA1607104, by the National Natural Science Foundation of China under Grants No. 12447105, No. 11775025, and No. 12175027.

\subsection*{Data Availability Statement}
The LZ recoil information and detector-efficiency input used in this analysis
are available in Ref.~\cite{LZ:2026axp}, including Fig.~S2 for the detector efficiency.
The numerical results are presented in the figures and in
Table~\ref{tab:benchmark_points}.

\bibliographystyle{apsrev4-1}
\raggedright
\bibliography{aps-longbibliography,ref}
\end{document}